\documentclass[12pt]{article}
\usepackage{graphicx}
\oddsidemargin=\evensidemargin
\begin{document}
\begin{center}
{\LARGE\bf Mass and Width of $\sigma(750)$ Scalar Meson}\\
{\LARGE\bf from Measurements of $\pi^- p \to \pi^- \pi^+ n$ on}\\
{\LARGE\bf Polarized Target$^*$}\\ 

\bigskip
\bigskip
{\Large Miloslav Svec}\\

\bigskip
\bigskip
{\it Department of Physics, McGill University}\\
{\it Montreal, PQ, Canada\ \ H3A 2T8}\\
\end{center} 

\bigskip
\bigskip
\bigskip
{\leftskip=2em\rightskip=2em\small
\noindent
{\bf Abstract.} Model independent amplitude analyses of $\pi^-
p_\uparrow \to \pi^- \pi^+ n$ on polarized target at 17.2 GeV/c reveal
resonant structure of $S$-wave transversity amplitudes $|\overline
S|^2\Sigma$ and $|S|^2\Sigma$ near 750 MeV. Simultaneous fit to
$|\overline S|^2\Sigma$ and $|S|^2\Sigma$ with a single $\sigma$ pole
yields $m_\sigma = 775 \pm 17$ MeV and $\Gamma_\sigma = 147 \pm 33$
MeV. Simultaneous fit with two common $\sigma$ poles yields a lower
$\chi^2/dpt$. Their resonance parameters are $m_\sigma = 786 \pm 24$
MeV, $\Gamma_\sigma = 130 \pm 47$ MeV and $m_{\sigma^\prime} = 670 \pm
30$ MeV, $\Gamma_{\sigma^\prime} = 59 \pm 58$ MeV.\par}
\bigskip
\medskip

In 1978, Lutz and Rybicki showed [1] that mesurements of pion
production in meson-nucleon scattering on transversely polarized target
allow model independent determination of production spin amplitudes.
This enables us to study resonance production on the level of
spin-dependent amplitudes rather than spin-averaged cross section
$\Sigma = d\sigma/dmdt$. Measurements of $\pi^- p_\uparrow\to
\pi^-\pi^+ n$ at 17.2 GeV/c [2], [3] and $\pi^+ n_\uparrow \to \pi^+
\pi^- p$ at 5.98 and 11.85 GeV/c [4] on polarized target were done at
CERN. The amplitude analysis provides mass distributions $|\overline
S|^2\Sigma$ and $|S|^2\Sigma$ of the two $S$-wave amplitudes $\overline
S$ and $S$ corresponding to nucleon transversity ``up'' and ``down'',
respectively. There are two solutions for each transversity amplitude.
 
The analysis of $\pi^- p_\uparrow \to \pi^- \pi^+ n$ data in the mass
range 600--900 MeV [2] shows that both solutions for $|\overline
S|^2\Sigma$ resonate around 750 MeV while $|S|^2\Sigma$ appears
nonresonating [5], [6]. The results indicate existence of a scalar
meson $\sigma(750)$ and are shown in Figure 1. The Breit-Wigner fit
with a coherent background to the resonating amplitude $|\overline
S|^2\Sigma$ yields $m_\sigma =753\pm 19$ MeV and $\Gamma_\sigma = 108
\pm 53$ MeV [6]. Further evidence for $\sigma(750)$ resonance comes
from measurements of $\pi^+ n_\uparrow \to \pi^+ \pi^- p$ at larger
momentum transfers. In [5], [6] we show that all four solutions to
$S$-wave intensity $I_S = (|S|^2 + |\overline S|^2)\Sigma$ in $\pi^+
n_\uparrow \to \pi^+ \pi^- p$, resonate around 750 MeV at both energies
measured.
\smallskip
\footnoterule
\smallskip
\begin{list}
{*}
\item {\footnotesize To appear in the Proceedings of the Seventh
International Conference on Hadron Spectroscopy, Hadron 97, Brookhaven
National Labortory, 25--30 August, 1997.\par}
\end{list}
 
We have also performed a new analysis [7] of $\pi^- p \to \pi^- \pi^+
n$ extending the mass range to 580--1080 MeV using data from [3]. Our
purpose was to study the role of interference of $f_0 (980)$ with
$\sigma(750)$ and to better determine the resonance parameters of
$\sigma(750)$ resonance from fits to both amplitudes $|\overline
S|^2\Sigma$ and $|S|^2\Sigma$. First we performed separate fits to
$|\overline S|^2\Sigma$ and $|S|^2\Sigma$ using Breit-Wigner
parametrization for $\sigma(750)$ and $f_0(980)$, and a constant
coherent background. We obtained an excellent fit a low average
$\chi^2/dpt$ ($\chi^2$ per data point) $=0.222$.

\begin{figure}[t]
\begin{center}
\includegraphics[height=116mm]{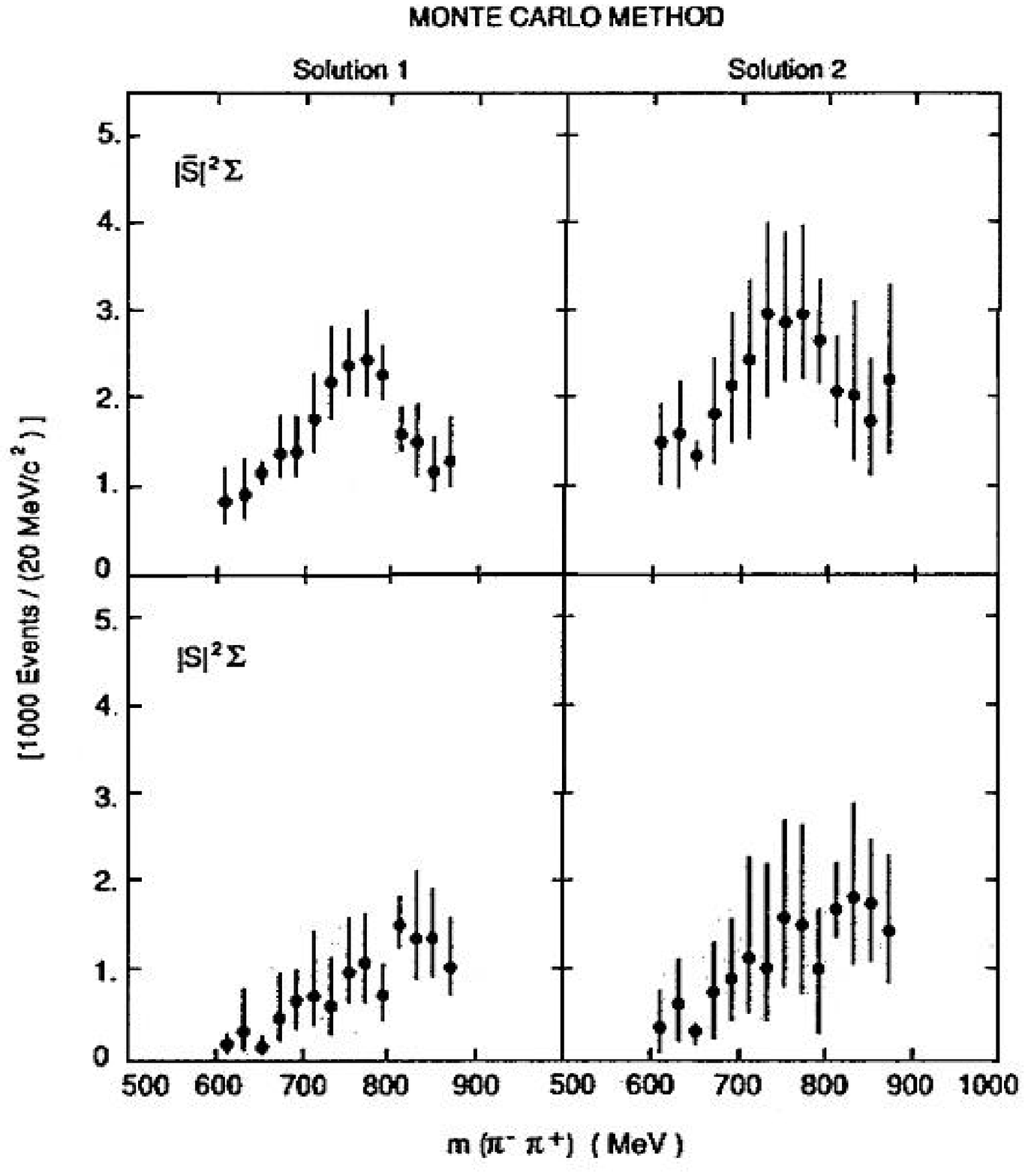}
\end{center}
\center{\small{\bf FIGURE 1.} Amplitudes $|\overline S|^2\Sigma$ and
$|S|^2\Sigma$ from data set [2].}
\end{figure}
However the width of $\sigma$ is narrow 76--110 MeV in $|\overline
S|^2\Sigma$ while it is broad 217--233 MeV in $|S|^2\Sigma$. To decide
whether the apparent dependence of $\Gamma_\sigma$ on nucleon spin is a
real effect or an artifact of separate fits, we performed simultaneous
fits to $|\overline S|^2\Sigma$ and $|S|^2\Sigma$. First we assumed a
common single $\sigma$ pole. We found a good fit with $m_\sigma =775\pm
17$ MeV and $\Gamma_\sigma = 147 \pm 33$ MeV but with a larger
$\chi^2/dpt = 0.330$. The results are shown in Figure 2.

Due to higher $\chi^2/dpt$ for simultaneous fit we concluded that the
results of separate fits may indicate existence of two $\sigma$ poles,
one with a narrow and the other with a broad width. Self-consistency
requires that both poles contribute to both $S$-wave amplitudes. The
simultaneous fit to $|\overline S|^2\Sigma$ and $|S|^2\Sigma$ with two
common $\sigma$ poles yields an excellent fit with average $\chi^2/dpt
= 0.181$.

All amplitudes are dominated by a $\sigma$ state with $m_\sigma =
786\pm 24$ MeV and $\Gamma_\sigma = 130\pm 47$ MeV.
\begin{figure}[t]
\begin{center}
\includegraphics[height=129mm]{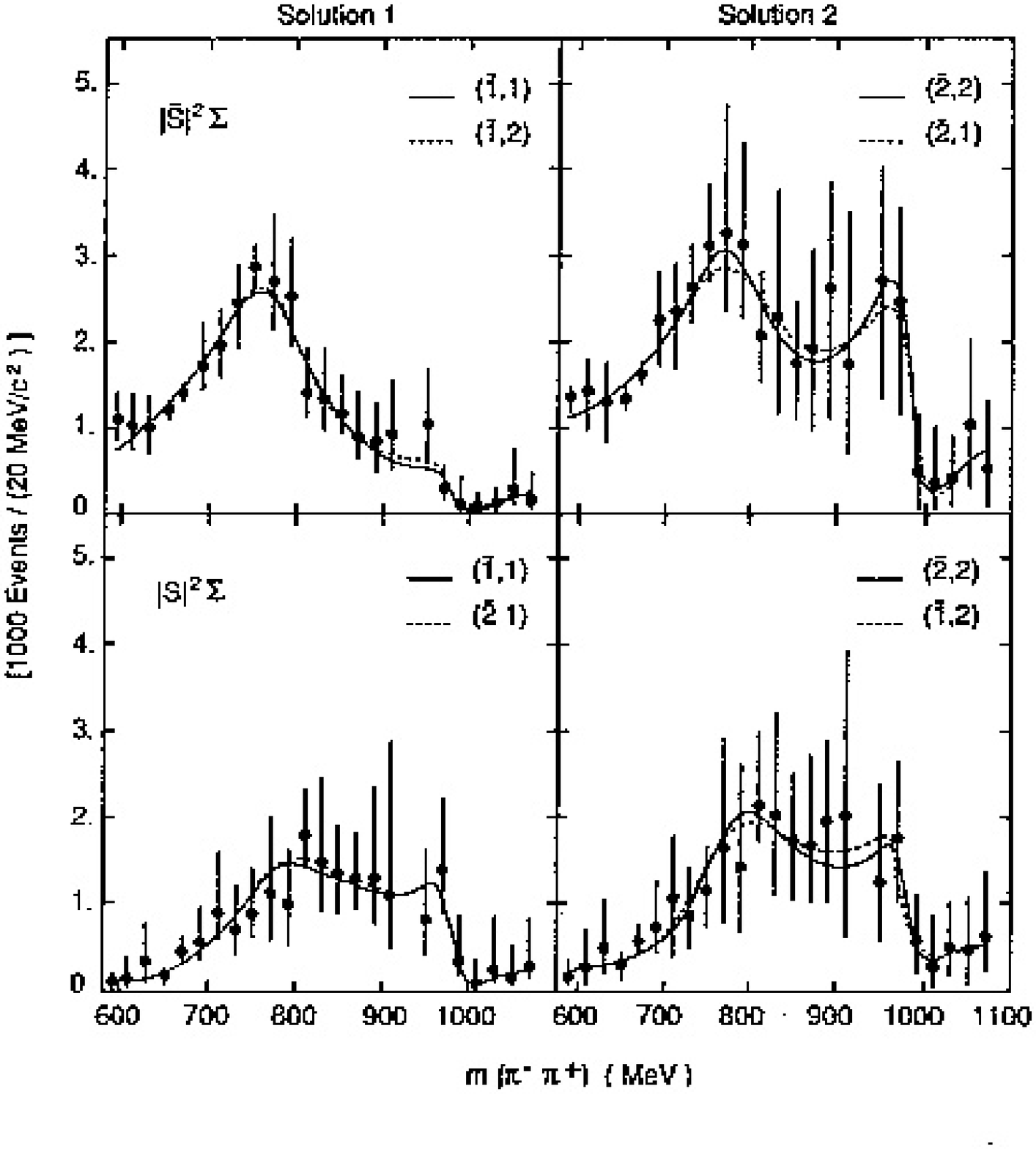}
\end{center}
\center{\small{\bf FIGURE 2.} Simultaneous fit with a
single $\sigma$ pole. Data from [3].}
\end{figure}
All amplitudes receive a weaker contribution from a narrow
$\sigma^\prime$ state with $m_{\sigma^\prime} = 670 \pm 30$ MeV and
$\Gamma_{\sigma^\prime} = 59 \pm 58$ MeV. The results are shown in
Figure 3. We propose to identify $\sigma^\prime(670)$ and $\sigma(786)$
with colour-electric and colour-magnetic modes of lowest mass scalar
gluonium $0^{++}(gg)$ [7].
\vspace*{10mm}
\begin{figure}[t]
\begin{center}
\includegraphics[height=129mm]{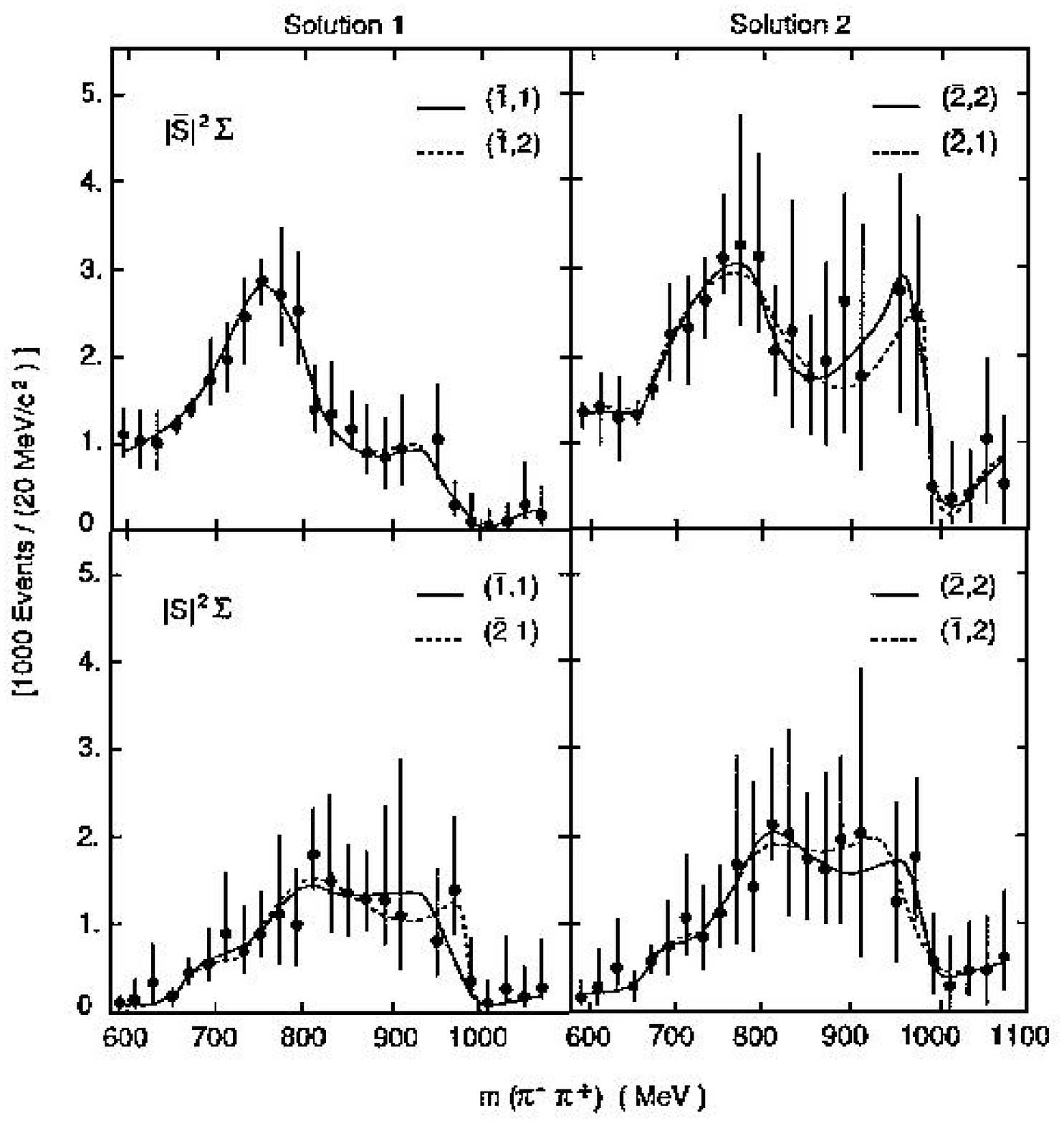}
\end{center}
\center{\small{\bf FIGURE 3.} Simultaneous fit with two $\sigma$ poles.
Data from [3].}
\end{figure}
\begin{center}
\large\bf
REFERENCES
\end{center}
\vspace{5mm}
{%
\parindent=0pt
G.~Lutz and K. Rybicki, Max Planck Institute, Munich, Internal Report
No.~MPS--PAE/Exp.~E1.75, 1978.

2. H.~Becker et al., Nucl.~Phys. \underbar{B150}, 301 (1979).

3. V.~Chabaud et al., Nucl.~Phys. \underbar{B223}, 1 (1983).

4. A.~de Lesquen et al., Phys.~Rev. \underbar{D32}, 21 (1985).

5. M.~Svec, Phys.~Rev. \underbar{D53}, 2343 (1996).

6. M.~Svec, Phys.~Rev. \underbar{D55}, 5727 (1997).

7. M.~Svec, McGill University preprint, 1997, hep--ph/9707495. 
}
\end{document}